\documentclass[11pt]{article}
\usepackage{amsmath,amsfonts,latexsym,graphicx,amssymb}
\usepackage{amsfonts}
\usepackage{bm}

\newcommand{\ot}{{\,\otimes\,}}
\newcommand{{\Cd}}{{\mathbb{C}^d}}

\def\<{\langle}
\def\>{\rangle}

\newtheorem{proposition}{Proposition}
\newtheorem{definition}{Definition}
\newtheorem{Example}{Example}
\newtheorem{Remark}{Remark}
\numberwithin{equation}{section}
\input{tcilatex}
\begin{document}

\date{}
\title{\textbf{On classical and quantum liftings}}
\author{L. Accardi$^1$, D. Chru\'sci\'nski$^2$, A. Kossakowski$^2$, T.
Matsuoka$^3$ and M. Ohya$^4$ \\
\\
$^1$ University of Roma Tor Vergata, Roma, Italy\\
$^2$ Nicolaus Copernicus University, Toru\'n, Poland\\
$^3$ Tokyo Universiy of Science, Suwa, Japan \\
$^4$ Tokyo Universiy of Science, Noda, Japan }
\maketitle

\begin{abstract}
We analyze the procedure of lifting in classical stochastic and quantum
systems. It enables one to `lift' a state of a system into a state of
`system+reservoir'. This procedure is important both in quantum information
theory and the theory of open systems. We illustrate the general theory of
liftings by a particular class related to so called circulant states.
\end{abstract}


\section{Introduction}

\setcounter{equation}{0}

The interest on quantum entanglement has dramatically increased
during the last two decades due to the emerging field of quantum
information theory \cite{QIT}. It turns out that quantum
entanglement may be used as {a basic resource} in quantum
information processing and communication. The prominent examples are
quantum cryptography, quantum teleportation, quantum
error correction codes and quantum computation (see the recent review \cite%
{HHHH}).

{The characteristic feature of an entangled state (already stressed
by Schr\"{o}dinger) is that knowing a composed system that is in a
pure state one can not say the same thing, in general, for any of
its subsystems. Therefore, a full knowledge of a system can
guarantee only a partial knowledge of its subsystems. This can't
happen in the classical case i.e. it is a purely quantum feature.}
In this paper we analyze {the converse} problem: suppose one knows a
state of a subsystem. How one can `lift' this state into a state of
the whole system? An appropriate mathematical framework was provided
in \cite{AO,BO}. This problem is important in many applications of
quantum information theory and quantum dynamics.

{For example it is of primary importance to be able to construct a
state of a composite system knowing only its marginals, i.e. the
states of subsystems.} An appropriate lifting enables one to define
a compound state which takes into account correlations between
subsystems \cite{AO,BO,BO1}.
Another important issue arises in the quantum dynamics of open systems \cite%
{Alicki,Breuer}. One is interested to extract {the dynamics of a
quantum system} from the unitary dynamics of the `system +
environment'. Due to the presence of system-environment correlations
this problem is highly nontrivial. One usually assumes that
initially `system + environment' was described by a product state,
i.e. there were no correlations at all. The standard {definition} of
reduced dynamics implies then that the dynamics of the system is
completely positive and trace preserving. A product state provides
{a very special} example of lifting: a system
state $\rho $ is lifted to `system + environment' state $\rho {\,\otimes \,}%
\omega $. Could we use other liftings to define initial `system +
environment' state?

One tries to relax the condition upon the `product lifting' and considers
dynamics which is only positive but not completely positive (see e.g. \cite%
{SS,D1,D2,Z}). Such positive dynamics enables one to map density matrices
into legitimate density matrices. However, one looses the important property
that coupling the system to an arbitrary environment will result in positive
`system + environment' dynamics. This problem appeared more recently in the
context of non-Markovian evolution (see recent papers \cite{NON-1,NON-2}).
An interesting debate upon this problem started already in the 90. of the
last century \cite{Pechukas,Alicki2}.

The problem of positive but not completely positive maps is very much
related to the problem of detection of quantum entanglement (see \cite{HHHH}
for the review). It is well known that the general structure of positive
maps is still unknown and only partial results are available (see \cite%
{P1,P2,P3,P4,P5,P6} for recent examples and constructions). One may try to
use appropriate liftings to provide new classes of such maps. Suppose for
example that we lift a state $\rho$ to the compound state \cite{AO,BO},
perform a positive map and then reduce to the system of interest. It is
obvious that the resulting map would be positive and it depends very much
upon the applied lifting. An interesting open problem is the classification
of `lifting assisted maps' defined this way.

The paper is organized as follows: for pedagogical reason we start with
presentation of basis ideas of stochastic classical systems Section \ref%
{CLASS}. In particular we show how to reformulate the classical theory in
the quantum framework. Section \ref{CLASS-LIFT} provides the discussion of
classical liftings and Section \ref{MULTI} provides generalization for
multipartite case. Then we consider the main topic of this paper in Section %
\ref{QUANT}. Section \ref{CIRC} illustrates the general ideas of liftings by
`circulant liftings'. These are liftings related to the construction of so
called circulant states \cite{CIRCULANT}. In particular we analyze Bell
diagonal liftings which play important role in quantum information theory
(see \cite{Bell-1,Bell-2} and recent paper \cite{Taka}). Final conclusions
are collected in the last Section.

\section{The quantum framework for classical probability with finite state
space}

\label{CLASS} \setcounter{equation}{0}

\subsection{Preliminaries}

Let us consider a classical $n$-level stochastic system. {The
primary object for elementary classical probability, i.e.
probability on a finite state space, is the space of elementary
events} $\Omega =\{\omega _{1},\ldots ,\omega _{n}\}$. The
corresponding space of states $S(\Omega )$ consists of probability
distributions over $\Omega $
\begin{equation}
S(\Omega ):=\left\{ p=(p_{1},\ldots ,p_{n})\ \Big|\ p_{i}\geq 0\ ,\ \
\sum_{i=1}^{n}p_{i}=1\ \right\} \ .
\end{equation}
It is clear that $S(\Omega )$ is convex and the corresponding set of
extremal points -- pure states -- reads as follows
\begin{equation}
S_{0}(\Omega ):=\left\{ p\in S(\Omega )\ \Big|\ p_{i}^{2}=p_{i}\ ,\
i=1,\ldots ,n\ \right\} \ .
\end{equation}%
Hence, a state $p\in S(\Omega )$ is pure iff $p_{i}=\delta _{ij}$ for some $%
j\in \{1,\ldots ,n\}$. It is evident that $S_{0}(\Omega )$ consists
of {the} $n$ vertices of $(n-1)$-dimensional simplex $S(\Omega )$. A
classical observable is a random variable $a\ :\ \Omega \
\longrightarrow \ \mathbb{R}\,$, and the corresponding expectation
value of $a$ in the state $p\in S(\Omega )$ is given by
\begin{equation}
\<a,p\>:=\sum_{i=1}^{n}a_{i}p_{i}=\sum_{i=1}^{n}a(\omega
_{i})P(\omega _{i})\ , \label{a-p}
\end{equation}%
where $a_{i}=a(\omega _{i})$ and $p_{i}=P(\omega _{i})$ (with $P:\Omega
\rightarrow \mathbb{R}$ being a probability distribution). Let us recall the
standard way of translating the above formulation of the classical
probability theory using the noncommutative framework of quantum theory.
Note, that the space of complex random variables
\begin{equation}
\psi \ :\ \Omega \ \longrightarrow \ \mathbb{C}\ ,
\end{equation}%
defines a Hilbert space
\begin{equation}
\mathcal{H}(\Omega ):=\left\{ \psi =(\psi _{1},\ldots ,\psi _{n})\ \Big|\
\psi _{i}=\psi (\omega _{i})\ \right\} \ \cong \ \mathbb{C}^{n}\ ,
\end{equation}%
equipped with the standard inner product
\begin{equation}
\<\varphi |\psi \>=\sum_{i=1}^{n}\,%
\overline{\varphi }_{i}\,\psi _{i}\ .
\end{equation}%
Now, any classical state $p\in S(\Omega )$ gives rise to a density matrix
living in $\mathcal{H}(\Omega )$
\begin{equation}
p=(p_{1},\ldots ,p_{n})\ \longrightarrow \ \rho
=\sum_{i=1}^{n}p_{i}\,e_{ii}\ ,  \label{p-rho}
\end{equation}%
where $e_{ii}=|i\>\<i|$ and $ e_{i}\equiv |i\> $ denotes an
arbitrary (but fixed) orthonormal basis in $\mathcal{H}(\Omega )$.
Hence classical states are represented by diagonal density matrices
with respect to the same fixed basis $e_{i}$. {Formula (\ref{p-rho})
defines therefore a decomposition of }$\rho $  which coincides with
the spectral decomposition when all the $p_{i}$'s are different.
Thus $S(\Omega )$ defines a commutative {subalgebra} of
\begin{equation}
S(\Omega )\ \subset \ S(\mathbb{C}^{n}):=\left\{ \ \rho :\mathbb{C}%
^{n}\longrightarrow \mathbb{C}^{n}\ |\ \rho \geq 0\ ,\ \mathrm{Tr}\,\rho =1\
\right\} \ ,
\end{equation}%
consisting of diagonal matrices. Classical pure states
\begin{equation}
S_{0}(\Omega )\ \subset \ S_{0}(\mathbb{C}^{n}):=\left\{ \ \rho \in S(%
\mathbb{C}^{n})\ |\ \rho ^{2}=\rho \ \right\} \ ,
\end{equation}%
are represented by {the} rank-1 projectors $e_{ii}$. In the same way
classical random variables define a commutative {subalgebra} of $%
\mathcal{B}(\mathbb{C}^{n})$:
\begin{equation}
a=(a_{1},\ldots ,a_{n})\ \longrightarrow \ a=\sum_{i=1}^{n}a_{i}\,e_{ii}\ ,
\label{a-a}
\end{equation}
that is, one may introduce the \textquotedblleft
classical\textquotedblright\ algebra $\mathcal{B}(\Omega )\subset \mathcal{B}%
(\mathbb{C}^{n})$
\begin{equation}
\mathcal{B}(\Omega ):=\left\{ \ a\in \mathcal{B}(\mathbb{C}^{n})\ \Big|\
a=\sum_{i=1}^{n}a_{i}\,e_{ii}\right\} \ .
\end{equation}%
An element $a\in \mathcal{B}(\Omega )$ defines a classical
observable iff $a^{\ast }=a$, i.e. $\overline{a_{i}}=a_{i}$.
Finally, the classical formula (\ref{a-p}) may be rewritten in a
\textquotedblleft quantum fashion\textquotedblright\ as follows
\begin{equation}
\< a,p \>=\mbox{Tr}\,a\rho \ . \label{a-rho}
\end{equation}

\subsection{Classical channels}

A classical channel is a linear positive map
\begin{equation}
\Lambda \ :\ \mathcal{B}(\Omega _{1})\ \longrightarrow \ \mathcal{B}(\Omega
_{2})\ ,
\end{equation}
where $\Omega _{1}=\{\omega _{1},\ldots ,\omega _{n_{1}}\}\,$ and
$\Omega =\{\varpi _{1},\ldots ,\varpi _{n_{2}}\}\,$. A channel is
unital iff $\Lambda (\mathbb{I}_{1})=\mathbb{I}_{2}$, where $
\mathbb{I}_{k}$ denotes an identity element in the $\mathbb{C}^{\ast
}$ -algebra $\mathcal{B}(\Omega _{k})$. Let
\begin{equation}
\mathcal{B}(\Omega _{1}):=\left\{ \ a\in \mathcal{B}(\mathbb{C}^{n_{1}})\ %
\Big|\ a=\sum_{i=1}^{n_{1}}a_{i}\,e_{ii}\right\} \ ,
\end{equation}
and
\begin{equation}
\mathcal{B}(\Omega _{2}):=\left\{ \ b\in \mathcal{B}(\mathbb{C}^{n_{2}})\ %
\Big|\ b=\sum_{i=1}^{n_{2}}b_{i}\,f_{ii}\right\} \ ,
\end{equation}
where $e_{i}$ defines an orthonormal basis in $\mathcal{H}(\Omega _{1})\cong
\mathbb{C}^{n_{1}}$, and $f_{i}$ defines an orthonormal basis in $\mathcal{H}%
(\Omega _{2})\cong \mathbb{C}^{n_{2}}$. One obtains
\begin{equation}
\Lambda (a)=\sum_{i=1}^{n_{1}}a_{i}\Lambda
(e_{ii})=\sum_{i=1}^{n_{1}}\sum_{j=1}^{n_{2}}a_{i}\Lambda _{ij}f_{jj}\ ,
\end{equation}
where
\begin{equation}
\Lambda _{ij}:=\mbox{Tr}(f_{jj}\Lambda (e_{ii}))\ .
\end{equation}
A linear map $\Lambda $ is positive if and only if $\Lambda _{ij}\geq 0$. A
channel $\Lambda $ transforms $a=\sum_{i}a_{i}e_{ii}$ into $%
b=\sum_{j}b_{j}f_{jj}$, with $b_{j}=\sum_{i}a_{i}\Lambda _{ij}$. It
is represented by {the} $n_{1}\times n_{2}$ matrix ($\Lambda _{ij})$
and it has the following Kraus representation
\begin{equation}
\Lambda (a)=\sum_{i=1}^{n_{1}}\sum_{j=1}^{n_{2}}\,K_{ij}\,a\,K_{ij}^{\ast }\
,  \label{K-K}
\end{equation}%
where the operators $K_{ij}:\mathbb{C}^{n_{1}}\longrightarrow \mathbb{C}%
^{n_{2}}$ are defined by
\begin{equation}
K_{ij}:=\sqrt{\Lambda _{ij}}\ |f_{j}\>\<e_{i}|\ , \label{K-ij}
\end{equation}
{and } $\sqrt{\Lambda _{ij}}$ is any complex valued square root of
$\left( \Lambda _{ij}\right) $. Let us observe that the above
representation of the classical channel involves operators which are
not diagonal. A channel is unital if
\begin{equation}
\sum_{i=1}^{n_{1}}\Lambda _{ij}=1\ ,\ \ \ \ j=1,\ldots ,n_{2}\ .
\end{equation}
It is therefore clear that a unital channel defines the conditional
probability
\begin{equation}
p_{i|j}:=\Lambda _{ij}\ ,
\end{equation}
on the space $\Omega _{1}\times \Omega _{2}$.

\begin{example}[Unitary channel]
\emph{Consider a channel
\begin{equation}
\Lambda \ :\ \mathcal{B}(\Omega )\ \longrightarrow \ \mathcal{B}(\Omega )\ ,
\end{equation}
defined by
\begin{equation}
\Lambda (a):=UaU^{\ast }\ ,
\end{equation}%
where $U$ is an $n\times n$ unitary matrix corresponding to a permutation $%
\pi $ from the symmetric group $S_{n}$, i.e.
\begin{equation}
U_{ij}:=\delta _{i\pi (j)}\ ,\ \ \ \ i,j=1,\ldots ,n\ .
\end{equation}%
One easily finds
\begin{equation}
\Lambda (a)=\sum_{i=1}^{n}a_{i}e_{\pi (i)\pi (i)}\ ,
\end{equation}%
where $\Lambda $ is represented by the following doubly stochastic matrix
\begin{equation}
\Lambda _{ij}=\mbox{Tr}(e_{jj}\Lambda (e_{ii}))=\delta _{j\,\pi (i)}\ .
\end{equation}%
Interestingly one has $\Lambda =U^{T}$. }
\end{example}

\begin{example}[Completely depolarizing channel]
\emph{Consider the unital channel
\begin{equation}
\Lambda \ :\ \mathcal{B}(\Omega _{1})\ \longrightarrow \ \mathcal{B}(\Omega
_{2})\ ,
\end{equation}%
defined by
\begin{equation}
\Lambda (a)=\frac{1}{n_{1}}\,\mathbb{I}_{2}\,\mathrm{Tr}\,a.
\end{equation}%
corresponding to the matrix $\left(\Lambda _{ij}\right)$ given by:
\begin{equation}
\Lambda _{ij}=\frac{1}{n_{1}}\ .
\end{equation}%
Note, that for $n_{1}=n_{2}$, the matrix $\left(\Lambda _{ij}\right)$ is doubly
stochastic. The corresponding representation (\ref{K-K}) reads as follows
\begin{equation}
\Lambda (a)=\sum_{i=1}^{n_{1}}\sum_{j=1}^{n_{2}}\,K_{ij}\,a\,K_{ij}^{\ast }\
,
\end{equation}%
where the $K_{ij}$ are defined by (\ref{K-ij}). }
\end{example}

\subsection{Composite classical systems}

Consider now a composite system consisting of two subsystems {with
state spaces} $\Omega _{1}$ and $\Omega _{2}$. The corresponding
space of elementary events has the form
\begin{equation}
\Omega _{12}\,=\,\Omega _{1}\times \Omega _{2}\ ,
\end{equation}%
hence $|\Omega _{12}|=n_{1}n_{2}$. The corresponding space of states
consists of the joint probability distributions $\,P:\Omega
_{12}\longrightarrow \mathbb{R}$
\begin{equation}
S(\Omega _{12}):=\left\{ p_{ij}\ \Big|\ p_{ij}\geq 0\ ,\ \
\sum_{i=1}^{n_{1}}\sum_{j=1}^{n_{2}}p_{ij}=1\ \right\} \ ,
\end{equation}%
where $p_{ij}=P(\omega _{i},\varpi _{j})$. Again, one may translate
multipartite case into the \textquotedblleft
quantum\textquotedblright\ framework by {considering the} complex
random variables $\psi \ :\ \Omega _{12}\ \longrightarrow \
\mathbb{C}\,$ {which} give rise to the following Hilbert space
\begin{equation}
\mathcal{H}(\Omega _{12})\,=\,\mathcal{H}(\Omega _{1})\otimes
\mathcal{H} (\Omega _{2})\,\cong \,\mathbb{C}^{n_{1}}{\,\otimes
\,}\mathbb{C}^{n_{2}}\ ,
\end{equation}
hence the classical space $S(\Omega _{12})$ defines a subspace of
$S( \mathcal{H}(\Omega _{12}))$
\begin{equation}
S(\Omega _{12})\,\subset \,S(\mathcal{H}(\Omega _{12})):=\left\{ \
\rho : \mathbb{C}^{n_{1}}{\,\otimes
\,}\mathbb{C}^{n_{2}}\longrightarrow \mathbb{C}
^{n_{1}}{\,\otimes \,}\mathbb{C}^{n_{2}}\ \Big|\ \rho \geq 0\ ,\ \mathrm{Tr}%
\,\rho =1\ \right\} \ ,
\end{equation}%
consisting of diagonal matrices
\begin{equation}
\rho \,=\,\sum_{i=1}^{n_{1}}\sum_{j=1}^{n_{2}}p_{ij}\,e_{ii}{\,\otimes \,}%
f_{jj}\ .
\end{equation}

\begin{remark}[A classical analogue of the maximally entangled state]
\emph{Consider a bipartite system with $\,\Omega _{12}=\Omega \times \Omega
\,$ and $\,|\Omega |=n$. For any $U\in U(n)$ one defines a maximally
entangled state in $\mathbb{C}^{n}{\,\otimes \,}\mathbb{C}^{n}$
\begin{equation}
P_{U}:=\frac{1}{n}\sum_{i,j=1}^{n}\,e_{ij}{\,\otimes \,}Ue_{ij}U^{\ast }\ .
\end{equation}%
In the classical case the unitaries correspond to permutations from the
symmetric group $S_{n}$ and one only considers diagonal states. Hence, the
classical counterpart of $P_{U}$ is given by
\begin{equation}
P_{\pi }=\frac{1}{n}\sum_{i=1}^{n}\,e_{ii}{\,\otimes \,}e_{\pi (i)\pi (i)}\ .
\end{equation}%
Note, that contrarily to $P_{U}$, its classical counterpart $P_{\pi }$ is no
longer a pure state. One may call $P_{\pi }$ \textit{a maximally correlated
classical state}.}
\end{remark}

\begin{remark}[A classical analog of the Choi-Jamio{\l }kowski isomorphism]
\emph{There is an evident isomorphism between classical channels and
classical states from $S(\Omega \times \Omega )$:} \emph{\ if
$\Lambda :\mathcal{B}(\Omega )\longrightarrow \mathcal{B}(\Omega )$
is a unital channel, then
\begin{equation}
\rho =\frac{1}{n}\,\sum_{i=1}^{n}\,e_{ii}{\,\otimes \,}\Lambda ^{\#
}(e_{ii})\ ,
\end{equation}%
where $\Lambda ^{\# }:\mathcal{S}(\Omega )\longrightarrow \mathcal{S}%
(\Omega )$ is the dual map, defines a classical state of the composite
system in $\Omega \times \Omega $. Note, that
\begin{equation}
\Lambda ^{\# }(e_{ii})=\sum_{j=1}^{n}p_{j|i}\,e_{jj}\ ,
\end{equation}%
implies
\begin{equation}
\rho =\sum_{i,j=1}^{n}p_{ij}\,e_{ii}{\,\otimes \,}e_{jj}\ ,
\end{equation}%
where $p_{ij}$ is the joint probability defined by $p_{ij}=p_{i|j}p_{j}^{(0)}
$, and $p^{(0)}$ corresponds to the uniform probability distribution $%
p_{k}^{(0)}=\frac{1}{n}$.}
\end{remark}

\begin{remark}[A classical analog of the teleportation protocol]
{The}\emph{\ state $P_{\pi }$ may be used to perform a classical
teleportation protocol: suppose that Alice would like to teleport a
classical state
\begin{equation}
\rho _{A}=\sum_{i=1}^{n}\,p_{i}\,e_{ii}\ .
\end{equation}%
In analogy to the quantum case one defines {the} joint state $\rho
_{A}{\,\otimes \,}P_{\pi }$ living in $\mathcal{H}(\Omega ){\,\otimes \,}%
\mathcal{H}(\Omega ){\,\otimes \,}\mathcal{H}(\Omega )$ and performs
 the joint measurement of $\,n^{2}P_{0}{\,\otimes
\,}\mathbb{I}_{B}\,$,
where $P_{0}$ corresponds to the trivial permutation, i.e. $%
P_{0}=n^{-1}\sum_{i}e_{ii}{\,\otimes \,}e_{ii}$, and $\mathbb{I}_{B}$ is an
identity operator in the Bob space. One obtains
\begin{equation}
\rho _{B}=n^{2}\,\mbox{Tr}_{12}(P_{0}{\,\otimes \,}\mathbb{I}_{B}\cdot \rho
_{A}{\,\otimes \,}P_{\pi })=\sum_{i=1}^{n}p_{\pi^{-1} (i)}e_{ii}\ .
\end{equation}%
It is therefore clear that performing {the} unitary transformation
corresponding to $\pi $ Bob recovers $\rho _{A}$. Summarizing: a
classical teleportation channel consists in simple permutation
$p_{i}\longrightarrow p_{\pi (i)}$, where the permutation $\pi $
corresponds to the `maximally entangled state' $P_{\pi }$ used in
the teleportation protocol. It is clear that there is a fundamental
difference between classical and quantum teleportation: in the
quantum case Bob needs 1 bit of additional classical information
from Alice. It is no longer needed in the classical case. Knowing
$\pi $ Bob reconstructs $\rho _{A}$ without any additional
information. }
\end{remark}

Consider a classical system living in $\Omega $ and let $\rho $ be a
classical state from $S(\Omega )$. To define a channel let $\sigma $ be a
fixed state of the \textit{ancilla} -- again from $\Omega $ -- and define
\begin{equation}
\Lambda ^{\# }(\rho )=\mbox{Tr}_{2}(U\rho {\,\otimes \,}\sigma
U^{\ast })\ ,  \label{channel-UU}
\end{equation}%
where $U$ is an $n^{2}\times n^{2}$ unitary matrix corresponding to \textbf{a%
} permutation from the symmetric group $S_{n^{2}}$. The simplest example
corresponds to $n=2$: let $\rho =\sum_{i}p_{i}e_{ii}$ and $\sigma
=\sum_{i}q_{i}e_{ii}$. One finds the following $2^{2}=4$ channels
\begin{equation}
p_{i}\ \longrightarrow \ \sum_{j=1}^{2}\Lambda _{ij}^{(k)}p_{j}\ ,
\end{equation}%
with
\begin{equation}
\Lambda ^{(1)}=\left(
\begin{array}{cc}
1 & 0 \\
0 & 1%
\end{array}%
\right) \ ,\ \ \ \ \Lambda ^{(2)}=\left(
\begin{array}{cc}
0 & 1 \\
1 & 0%
\end{array}%
\right) \ ,
\end{equation}%
and
\begin{equation}
\Lambda ^{(3)}=\left(
\begin{array}{cc}
q_{1} & q_{2} \\
q_{2} & q_{1}%
\end{array}%
\right) =q_{1}\Lambda ^{(1)}+q_{2}\Lambda ^{(2)}\ ,\ \ \ \ \Lambda
^{(4)}=\left(
\begin{array}{cc}
q_{2} & q_{1} \\
q_{1} & q_{2}%
\end{array}%
\right) =q_{1}\Lambda ^{(2)}+q_{2}\Lambda ^{(1)}\ .
\end{equation}%
Note that all 4 matrices $[\Lambda _{ij}^{(k)}]$ are doubly
stochastic. One easily proves the following general {statement.}

\begin{proposition}
A classical channel defined via the reduction procedure (\ref{channel-UU})
is doubly stochastic.
\end{proposition}

We stress that a quantum channel defined by the reduction procedure is trace
preserving but not necessarily unital.

\subsection{Multipartite case}

{The} bipartite case may be easily generalized to many parties.
Consider a composite system consisting of $N$ parties with
\begin{equation}
\Omega _{N\ldots 1}\,=\,\Omega _{N}\times \ldots \times \Omega _{1}\ ,
\end{equation}%
and $|\Omega _{k}|=n_{k}$. It implies $|\Omega _{1\ldots N}|=n_{1}\ldots
n_{N}$. The corresponding space of states consists of the joint probability
distributions $\,P:\Omega _{N\ldots 1}\longrightarrow \mathbb{R}$
\begin{equation}
S(\Omega _{N\ldots 1}):=\Big\{p_{i_{N}\ldots i_{1}}\ \Big|\ p_{i_{N}\ldots
i_{1}}\geq 0\ ,\ \ \sum_{i_{1}\ldots i_{N}}p_{i_{N}\ldots i_{1}}=1\ \Big\}\ ,
\end{equation}%
where $p_{i_{N}\ldots i_{1}}=P(\omega _{i_{N}}^{(N)},\ldots ,\omega
_{i_{1}}^{(1)})$, and $\omega _{i}^{(k)}$ is an elementary event from $%
\Omega _{k}$.

Again, one may translate multipartite case into the quantum framework. Note
that complex random variables
\begin{equation}
\psi\ : \ \Omega_{N\ldots 1}\ \longrightarrow\ \mathbb{C}\ ,
\end{equation}
give rise to the multipartite Hilbert space
\begin{equation}
\mathcal{H}(\Omega_{N\ldots 1})\, =\, \mathcal{H}(\Omega_N) \otimes \ldots
\otimes \mathcal{H}(\Omega_1) \, \cong\, \mathbb{C}^{n_N} {\,\otimes\,}
\ldots {\,\otimes\,} \mathbb{C}^{n_1}\ ,
\end{equation}
and hence the classical space $S(\Omega_{N\ldots 1})$ defines a subspace of $%
S(\mathcal{H}(\Omega_{N\ldots 1}))$ consisting of the diagonal density
matrices
\begin{equation}  \label{N-class}
\rho\, =\, \sum_{i_1=1}^{n_1}\ldots \sum_{i_N=1}^{n_N} p_{i_N\ldots i_1} \,
e^{(N)}_{i_Ni_N}{\,\otimes\,} \ldots {\,\otimes\,} e^{(1)}_{i_1i_1}\ ,
\end{equation}
where $\{e^{(k)}_{1},\ldots, e^{(k)}_{n_k}\}$ defines an orthonormal basis
in $\mathcal{H}(\Omega_k)$.

\subsection{Markovian states}

Consider now the special case when all subsystems live on the same space $%
\Omega $. A classical $N$-partite state on $\Omega \times \ldots \times
\Omega $
\begin{equation}
\rho =\sum_{i_{N},\ldots ,i_{1}=1}^{n}\,p_{i_{N}\ldots i_{1}}\,e_{i_{N}i_{N}}%
{\,\otimes \,}\ldots {\,\otimes \,}e_{i_{1}i_{1}}\ ,
\end{equation}%
is Markovian, if the $N$-partite joint probabilities $%
p_{i_{N}\ldots i_{1}}$ {having the following} property
\begin{equation}
p_{i_{N}\ldots i_{1}}=p_{i_{N}|i_{N-1}}\,p_{i_{N-1}|i_{N-2}}\,\ldots
\,p_{i_{2}|i_{1}}\,p_{i_{1}}\ ,  \label{p-Markov}
\end{equation}%
where $p_{i|j}$ defines the conditional probability on $\Omega
\times \Omega $ and $p_{i}$ {is a} probability distribution on
$\Omega $. If $\rho $
is Markovian, then its reduction with respect to {the} subsystems $%
N,N-1,\ldots ,N-k+1$ gives $(N-k)$-partite Markovian state
\begin{equation}
\mbox{ Tr}_{N}\ldots \mbox{ Tr}_{N-k+1}\,\rho \,=\,\sum_{i_{N-k},\ldots
,i_{1}=1}^{n}\,p_{i_{N-k}|i_{N-k-1}}\,\ldots
\,p_{i_{2}|i_{1}}\,p_{i_{1}}\,e_{i_{N-k}i_{N-k}}{\,\otimes \,}\ldots {%
\,\otimes \,}e_{i_{1}i_{1}}\ .  \label{N-class}
\end{equation}%
Let us translate the Markovian property into the quantum noncommutative
setting. Let $\mathcal{A}_{1}$ and $\mathcal{A}_{2}$ be $\mathbb{C}^{\ast }$%
-algebras.

\begin{definition}
A linear completely positive map
\begin{equation}
\mathcal{E}\ :\ \mathcal{A}_2 {\,\otimes\,} \mathcal{A}_1 \longrightarrow
\mathcal{A}_1\ ,
\end{equation}
is called a transition expectation if
\begin{equation}
\mathcal{E}({\mathchoice{\rm 1\mskip-4mu l}{\rm 1\mskip-4mu l}{\rm
1\mskip-4.5mu l}{\rm 1\mskip-5mu l}}_2 {\,\otimes\,} {\mathchoice{\rm
1\mskip-4mu l}{\rm 1\mskip-4mu l}{\rm 1\mskip-4.5mu l}{\rm 1\mskip-5mu l}}%
_1) = {\mathchoice{\rm 1\mskip-4mu l}{\rm 1\mskip-4mu l}{\rm 1\mskip-4.5mu
l}{\rm 1\mskip-5mu l}}_1\ ,
\end{equation}
where ${%
\mathchoice{\rm 1\mskip-4mu l}{\rm 1\mskip-4mu l}{\rm 1\mskip-4.5mu
l}{\rm 1\mskip-5mu l}}_k$ denotes an identity element in $\mathcal{A}_k$.
\end{definition}
One has the following

\begin{proposition}
An $N$-partite state $\rho$ on $\Omega^{ N}$ is Markovian if there exist a
transition expectation
\begin{equation}
\mathcal{E}\ :\ \mathcal{B}(\Omega) {\,\otimes\,} \mathcal{B}(\Omega)
\longrightarrow \mathcal{B}(\Omega)\ ,
\end{equation}
such that
\begin{equation}  \label{EEE}
\mbox{Tr} \left( \rho \, (a_N {\,\otimes\,} \ldots {\,\otimes\,} a_1)
\right) = \mbox{Tr} \left( \rho_1\, \mathcal{E}(a_N {\,\otimes\,} \mathcal{E}%
(a_{N-1} {\,\otimes\,} \ldots {\,\otimes\,} \mathcal{E}(a_1 {\,\otimes\,}
\mathbb{I})) \ldots ) \right) \ ,
\end{equation}
is satisfied for arbitrary $a_1,\ldots,a_N \in \mathcal{B}(\Omega)$, where $%
\rho_1 = \mbox{Tr}_{\breve{1}}\rho$ is the reduced density matrix living on
the first subsystem.
\end{proposition}
Indeed, introducing so called Markov operator $P:\mathcal{B}(\Omega
)\longrightarrow \mathcal{B}(\Omega )$
\begin{equation}
P(e_{ii})=\sum_{k=1}^{n}p_{i|j}\,e_{jj}\ ,
\end{equation}%
where $p_{i|j}$ denotes conditional probability, and defining {the}
transition expectation via
\begin{equation}
\mathcal{E}(a_{2}{\,\otimes \,}a_{1})=P(a_{2})a_{1}\ ,
\end{equation}%
one shows equivalence of (\ref{EEE}) and (\ref{p-Markov}).

\section{Classical liftings}

\label{CLASS-LIFT} \setcounter{equation}{0}

Let $\mathcal{A}_1$ and $\mathcal{A}_2$ be $\mathbb{C}^*$-algebras. One
introduces \cite{AO} the following

\begin{definition}  A lifting from $\mathcal{A}_1$ to
$\mathcal{A}_2 {\,\otimes\,} \mathcal{A}_1$ is a map
\begin{equation}  \label{1>2}
\mathcal{E}^\# \ :\ \mathcal{S}(\mathcal{A}_1)\ \longrightarrow\
\mathcal{S}(\mathcal{A}_2{\,\otimes\,} \mathcal{A}_1)\ .
\end{equation}
If the map $\mathcal{E}^\#$ is affine and  its dual $\mathcal{E} :
\mathcal{A}_2{\,\otimes\,} \mathcal{A}_1 \longrightarrow
\mathcal{A}_1$ is completely positive, then one calls
$\mathcal{E}^\#$ a linear lifting. If $\mathcal{E}^\#$ maps pure
states into pure states, one calls it a pure lifting.
\end{definition}
Following \cite{AO} let us consider a lifting
\begin{equation}
\mathcal{E}^{\# }\ :\ \mathcal{S}(\Omega _{1})\ \longrightarrow \
\mathcal{S}(\Omega _{2}\times \Omega _{1})\ .  \label{L-class}
\end{equation}%
Note, that any linear lifting has the following form
\begin{equation}
\mathcal{E}^{\# }(\rho )=\sum_{i,j=1}^{n_{1}}\sum_{k=1}^{n_{2}}\,p_{i}%
\mathcal{E}_{ijk}\,f_{jj}{\,\otimes \,}e_{kk}\ ,
\end{equation}%
where
\begin{equation}
\mathcal{E}_{ijk}:=\mbox{Tr}\left( f_{jj}{\,\otimes \,}e_{kk}\,\mathcal{E}%
^{\# }(e_{ii})\right) =\mbox{Tr}\left( e_{ii}\,\mathcal{E}(f_{jj}{%
\,\otimes \,}e_{kk})\right) \ ,
\end{equation}%
and $\rho =\sum_{i}p_{i}e_{ii}$. One has
\begin{equation}
\mathcal{E}^{\# }(\rho
)=\sum_{j=1}^{n_{2}}\sum_{k=1}^{n_{1}}\,p_{jk}\,f_{jj}{\,\otimes
\,}e_{kk}\ ,
\end{equation}%
and hence
\begin{equation}
p_{jk}=\sum_{i=1}^{n_{1}}\,\mathcal{E}_{ijk}\,p_{i}\ .
\end{equation}

\begin{definition}
A lifting $\mathcal{E}^\#$ is called non-demolishing for a state
$\rho \in \mathcal{S}(\Omega_1)$ if
\begin{equation}
\mathrm{Tr}_2 \,\mathcal{E}^\#(\rho) = \rho\ .
\end{equation}
\end{definition}
It is clear that if $\mathcal{E}^{\# }$ is linear and
non-demolishing for some $\rho $, then it is non-demolishing for all
states. In this case one has
\begin{equation}
\sum_{k=1}^{n_{2}}\mathcal{E}_{ijk}=\delta _{ij}\ .
\end{equation}

\begin{remark} {\em The notion of nondemolition lifting defined
here is essentially (i.e., up to minor technicalities) included in
the more abstract notion of state extension introduced by Cecchini
and Petz \cite{CP1} (see also \cite{CP2}).}
\end{remark}

\begin{definition}
A lifting $\mathcal{E}^{\# }$ is Markovian lifting if
\begin{equation}
\mathcal{E}^{\#
}(e_{ii})=\sum_{j=1}^{n_{2}}\,p_{j|i}\,f_{jj}{\,\otimes \,}e_{ii}\ ,
\end{equation}%
where $p_{j|i}$ stands for the conditional probability on $\Omega _{2}\times
\Omega _{1}$.
\end{definition}

\begin{example}[Pure lifting]
\emph{Let $s:\{1,\ldots ,n_{1}\}\longrightarrow \{1,\ldots ,n_{2}\}$ be a
function and define
\begin{equation}
\mathcal{E}^{\# }(e_{ii})=f_{s(i)s(i)}{\,\otimes \,}e_{ii}\ ,
\end{equation}%
that is, it lifts a pure state on $\Omega _{1}$ into pure states on $\Omega
_{2}\times \Omega _{1}$. Hence, for $\rho =\sum_{i}p_{i}\,e_{ii}$ one finds
\begin{equation}
\mathcal{E}^{\# }(\rho
)=\sum_{i=1}^{n_{1}}p_{i}\,f_{s(i)s(i)}{\,\otimes \,}e_{ii}\ ,
\end{equation}%
which implies
\begin{equation}
\mathcal{E}_{ijk}=\delta _{ik}\delta _{j,s(i)}\ .
\end{equation}%
}
\end{example}

\begin{Example}[Product lifting]
\emph{Let
\begin{equation}
\mathcal{E}_{ijk} = \delta_{ik}\, q_j \ ,
\end{equation}
where $\sigma = \sum_k q_k f_{kk}$ is a classical state over $\Omega_2$. One
obtains
\begin{equation}
\mathcal{E}^\#(\rho) = \sigma {\,\otimes\,} \rho \ ,
\end{equation}
that is, a joint probability is given by a product formula $p_{jk} = p_j q_k$%
. }
\end{Example}

\begin{Example}
\emph{Let $\sigma$ be a fixed state on $\Omega_2$ and
\begin{equation}
\Gamma \ : \ \mathcal{B}(\Omega_2 \times \Omega_1) \longrightarrow \mathcal{B%
}(\Omega_2 \times \Omega_1) \ ,
\end{equation}
be a linear positive map. Define the lifting $\mathcal{E}^{\#\Gamma,%
\sigma}$ by
\begin{equation}
\mathcal{E}^{\#\,\Gamma,\sigma}(\rho) := \Gamma^\#(\sigma{\,\otimes\,%
} \rho)\ .
\end{equation}
Note, that if $\Gamma = \mathrm{id}_2 {\,\otimes\,} \mathrm{id}_1$, then the
above formula recovers a product lifting. }
\end{Example}

\begin{Example}[Classical Ohya lifting]
\emph{If
\begin{equation}
\mathcal{E}_{ijk} = \delta_{ik}\, \delta_{jk} \ ,
\end{equation}
then
\begin{equation}
\mathcal{E}^\#(\rho) = \sum_{i=1}^n p_i\, e_{ii} {\,\otimes\,}
e_{ii}\ .
\end{equation}
Note, that a characteristic feature of the Ohya lifting is that both reduced
states reproduce the original state $\rho$, i.e.
\begin{equation}
\mathrm{Tr}_1\,\mathcal{E}^\#(\rho) = \mathrm{Tr}_2\,\mathcal{E}%
^\#(\rho) = \rho\ ,
\end{equation}
for an arbitrary classical state $\rho$ on $\Omega$. It shows that Ohya
lifting realizes a perfect classical cloning machine \cite{clon}. Moreover,
since both reduced states are the same Ohya lifting provides at the same
time a protocol for a classical broadcasting \cite{Ariano}. }
\end{Example}

\section{Multipartite classical liftings}

\label{MULTI} \setcounter{equation}{0}

It is clear that a lifting (\ref{1>2}) may be easily generalized \textbf{to}
the multipartite case: consider a set $\mathcal{A}_{1},\ldots ,\mathcal{A}%
_{N}$ of $\mathbb{C}^{\ast }$-algebras.

\begin{definition}
An $N$-lifting from $\mathcal{A}_1$ to $\mathcal{A}_N {\,\otimes\,} \ldots {%
\,\otimes\,} \mathcal{A}_1$ is a map
\begin{equation}  \label{1>N}
\mathcal{E}^\# \ : \ \mathcal{S}(\mathcal{A}_1)\ \longrightarrow\
\mathcal{S}(\mathcal{A}_N{\,\otimes\,} \ldots {\,\otimes\,}
\mathcal{A}_1)\ .
\end{equation}
If the map $\mathcal{E}^\#$ is affine and its dual $\,\mathcal{E} :
\mathcal{A}_N{\,\otimes\,} \ldots {\,\otimes\,} \mathcal{A}_1
\longrightarrow \mathcal{A}_1$ is completely positive, then one calls $%
\mathcal{E}^\#$ a linear lifting.
\end{definition}
%
Applying the above definition to the classical commutative case one finds
\begin{equation}  \label{L-class-N}
\mathcal{E}_N^\# \ : \ \mathcal{S}(\Omega_1) \ \longrightarrow \
\mathcal{S}(\Omega_N \times \ldots \times \Omega_1)\ .
\end{equation}
Note, that any linear classical $N$-lifting has the following form
\begin{equation}
\mathcal{E}_N^\#(\rho) = \sum_{i=1}^{n_1}\sum_{i_1=1}^{n_1}\,
\ldots\,\sum_{i_N=1}^{n_N}\, \mathcal{E}_{i_N\ldots i_1 i}\, p_i \,
e^{(N)}_{i_Ni_N} {\,\otimes\,} \ldots {\,\otimes\,}
e^{(1)}_{i_1i_1}\ ,
\end{equation}
where
\begin{equation}
\mathcal{E}_{i_N\ldots i_1 i} := \mbox{Tr} \left( e^{(N)}_{i_Ni_N} {%
\,\otimes\,} \ldots {\,\otimes\,} e^{(1)}_{i_1i_1}\, \mathcal{E}%
_N^\#(e_{ii}^{(1)}) \right) = \mbox{Tr} \left( \mathcal{E}%
_N(e^{(N)}_{i_Ni_N} {\,\otimes\,} \ldots {\,\otimes\,} e^{(1)}_{i_1i_1})\,
e_{ii}^{(1)} \right)\ ,
\end{equation}
and $\rho = \sum_i p_i e_{ii}$. Note, that $\mathcal{E}_N^\#(\rho)$
is an $N$-partite classical state and hence it corresponds to an
$N$-partite joint probability distribution $p_{i_n\ldots i_1}$, that
is
\begin{equation}
\mathcal{E}_N^\#(\rho) = \sum_{i_1=1}^{n_1}\,
\ldots\,\sum_{i_N=1}^{n_N}\, p_{i_N\ldots i_1} \, e^{(N)}_{i_Ni_N} {%
\,\otimes\,} \ldots {\,\otimes\,} e^{(1)}_{i_1i_1}\ ,
\end{equation}
and hence
\begin{equation}
p_{i_N\ldots i_1} = \sum_{i_1=1}^{n_1}\, \mathcal{E}_{i_N\ldots i_1 i}\, p_i
\ .
\end{equation}

\begin{example}[Pure $N$-lifting]
\emph{\ Let $s_{k}:\{1,\ldots ,n_{1}\}\longrightarrow \{1,\ldots ,n_{k}\}$ ($%
k=2,3,\ldots ,N$) be a family of functions and define the following linear $N
$-lifting
\begin{equation}
\mathcal{E}_{N}^{\#
}(e_{ii}^{(1)})=e_{s_{N}(i)s_{N}(i)}^{(N)}{\,\otimes
\,}\ldots {\,\otimes \,}e_{s_{2}(i)s_{2}(i)}^{(2)}{\,\otimes \,}%
e_{ii}^{(1)}\ ,
\end{equation}%
that is, it lifts a pure state on $\Omega _{1}$ into pure states on $\Omega
_{N}\times \ldots \times \Omega _{1}$. Hence, for $\rho
=\sum_{i}p_{i}\,e_{ii}^{(1)}$ one finds
\begin{equation}
\mathcal{E}_{N}^{\# }(\rho
)=\sum_{i=1}^{n_{1}}p_{i}\,e_{s_{N}(i)s_{N}(i)}^{(N)}{\,\otimes \,}\ldots {%
\,\otimes \,}e_{s_{2}(i)s_{2}(i)}^{(2)}{\,\otimes \,}e_{ii}^{(1)}\ ,
\end{equation}%
which implies
\begin{equation}
\mathcal{E}_{i_{N}\ldots i_{1}i}=\delta _{i_{N},s_{N}(i)}\ldots \delta
_{i_{2},s_{2}(i)}\delta _{i_{1}i}\ .
\end{equation}%
}
\end{example}

\begin{Example}[Product $N$-lifting]
\emph{Let $\sigma_N {\,\otimes\,} \ldots {\,\otimes\,} \sigma_2$ be a
product state on $\Omega_N \times \ldots \times \Omega_2$. One defines
\begin{equation}
\mathcal{E}_N^\#(\rho) = \sigma_N {\,\otimes\,} \ldots {\,\otimes\,}
\sigma_2 {\,\otimes\,} \rho \ ,
\end{equation}
that is,
\begin{equation}
\mathcal{E}_{i_N\ldots i_1 i} = q^{(N)}_{i_N} \ldots q^{(2)}_{i_2}\,
\delta_{i_1i}\ ,
\end{equation}
where $\sigma_k = \sum_{i_k} q^{(k)}_{i_k} e^{(k)}_{i_ki_k}$. It is clear
that the corresponding joint probability factorizes as follows
\begin{equation}
p_{i_N\ldots i_1} = q^{(N)}_{i_N} \ldots q^{(2)}_{i_2}\, p_{i_1}\ .
\end{equation}
}
\end{Example}

\begin{Example}
\emph{Let $\sigma=\sigma_N {\,\otimes\,} \ldots {\,\otimes\,} \sigma_2$ be a
product state on $\Omega_N \times \ldots \times \Omega_2$ and
\begin{equation}
\Gamma \ : \ \mathcal{B}(\Omega_N \times \ldots \times \Omega_1)
\longrightarrow \mathcal{B}(\Omega_N \times \ldots \times \Omega_1) \ ,
\end{equation}
be a linear positive map. Define the $N$-lifting $\mathcal{E}%
^{\#\Gamma,\sigma}$ by
\begin{equation}
\mathcal{E}_N^{\#\,\Gamma,\sigma}(\rho) := \Gamma^\#(\sigma {%
\,\otimes\,} \rho)\ .
\end{equation}
Note, that if $\Gamma = \mathrm{id}_N {\,\otimes\,} \ldots {\,\otimes\,}
\mathrm{id}_1$, then the above formula recovers a product $N$-lifting. }
\end{Example}

\begin{Example}[Classical Ohya $N$-lifting]
\emph{Let $\Omega_1 = \ldots = \Omega_N = \Omega$. Taking
\begin{equation}
\mathcal{E}_{i_N\ldots i_1 i} = \delta_{i_Ni}\ldots \delta_{i_1i} \ ,
\end{equation}
one obtains
\begin{equation}
\mathcal{E}_N^\#(e_{ii}) = e_{ii} {\,\otimes\,} \ldots {\,\otimes\,}
e_{ii}\ ,
\end{equation}
and hence for $\rho =\sum_i p_i\, e_{ii}$ one finds
\begin{equation}
\mathcal{E}_N^\#(\rho) = \sum_{i=1}^n p_i\, e_{ii} {\,\otimes\,} \ldots {%
\,\otimes\,} e_{ii}\ .
\end{equation}
Note, that a characteristic feature of the Ohya lifting is that all reduced
one-partite states reproduce the original state $\rho$, i.e.
\begin{equation}
\mathrm{Tr}_{\breve{1}}\,\mathcal{E}^\#(\rho) = \ldots = \mathrm{Tr}_{%
\breve{N}}\,\mathcal{E}^\#(\rho) = \rho\ ,
\end{equation}
where $\mathrm{Tr}_{\breve{k}}$ denotes the partial trace with respect all
subsystems excluding $k$th. It shows that Ohya lifting realizes a perfect
classical cloning machine \cite{clon}. Moreover, since both reduced states
are the same Ohya lifting provides at the same time a protocol for a
classical broadcasting \cite{Ariano}. }
\end{Example}

\begin{Remark}
\emph{Let $\varphi_k : \mathcal{B}(\mathbb{C}^n) \longrightarrow \mathcal{B}(%
\mathbb{C}^n)$ be a set of positive unital maps (for $k=1,\ldots,N$). Then
the following $N$-partite state
\begin{equation}
\rho = \sum_{i=1}^n p_i\, \varphi^\#_1(e_{ii}) {\,\otimes\,} \ldots {%
\,\otimes\,} \varphi^\#_N(e_{ii})\ ,
\end{equation}
is a (quantum) $N$-separable state. }
\end{Remark}

Let us observe that in the special case when $\Omega_2 = \ldots = \Omega_N
=: \Omega$ there is a simple way to construct an $N$-lifting out of the
lifting
\begin{equation}  \label{L-class}
\mathcal{E}^\# \ : \ \mathcal{S}(\Omega_1) \ \longrightarrow \ \mathcal{S%
}(\Omega_2 \times \Omega_1)\ ,
\end{equation}
Define the $N$-lifting
\begin{equation}  \label{L-class-N}
\mathcal{E}_N^\# \ : \ \mathcal{S}(\Omega_1) \ \longrightarrow \
\mathcal{S}(\Omega_N \times \ldots \times \Omega_1)\ ,
\end{equation}
by the following recurrence formula: for $N=3$
\begin{equation}
\mathcal{E}_3^\# \ : \ \mathcal{S}(\Omega_1) \ \longrightarrow \
\mathcal{S}(\Omega_3 \times \Omega_2 \times \Omega_1)\ ,
\end{equation}
one has
\begin{equation}
\mathcal{E}_3^\# = (\mbox{id}_2 {\,\otimes\,} \mathcal{E}^\#) \circ
\mathcal{E}^\# \ ,
\end{equation}
and for $N > 3 $
\begin{equation}
\mathcal{E}_N^\# = (\mathrm{id}_{N-1} {\,\otimes\,} \ldots
{\,\otimes\,}
\mathrm{id}_2 {\,\otimes\,} \mathcal{E}^\#) \circ \mathcal{E}%
_{N-1}^\# \ .
\end{equation}
An example of this construction is provided by the Ohya $N$-lifting.

\section{Quantum noncommutative liftings}

\label{QUANT} \setcounter{equation}{0}

\subsection{Linear liftings}

To any transition expectation represented by a completely positive linear
unital map

\begin{equation}
\mathcal{E}\ :\ \mathcal{A}_{2}{\,\otimes \,}\mathcal{A}_{1}\
\longrightarrow \ \mathcal{A}_{1}\ ,
\end{equation}%
there corresponds a linear lifting
\begin{equation}
\mathcal{E}^{\# }\ :\ S(\mathcal{A}_{1})\ \longrightarrow \ S(\mathcal{A}%
_{2}{\,\otimes \,}\mathcal{A}_{1})\ .
\end{equation}%
Linear liftings play important role in {the} quantum theory of open
systems. It is well known that a linear lifting which is
nondemolishing for all states $\rho $ has the following form
\begin{equation}
\mathcal{E}^{\# }(\rho )=\omega {\,\otimes \,}\rho \ ,  \label{x}
\end{equation}%
for some fixed state $\omega $ over $\mathcal{A}_{2}$. Let us assume that $%
\mathcal{A}_{k}=\mathcal{B}(\mathcal{H}_{k})$. A product lifting (or
assignment maps \cite{Pechukas,Alicki}) defined by (\ref{x}) gives rise to a
linear completely positive map
\begin{equation*}
\Lambda ^{\# }\ :\ S(\mathcal{A}_{1})\ \longrightarrow \ S(\mathcal{A}%
_{1})\ ,
\end{equation*}%
defined by
\begin{equation}
\Lambda ^{\# }(\rho ):=\mathrm{Tr}_{2}[\,U(\omega {\,\otimes \,}\rho
)U^{\ast }\,]\ ,
\end{equation}%
where $U$ is a unitary operator in $\mathcal{H}_{2}{\,\otimes \,}\mathcal{H}%
_{1}$. It is well known that any linear completely positive map may be
obtained this way.

\begin{remark} {\em Let us observe that linear lifting may be used
to construct new classes of positive maps which are not completely
positive.  Let $\mathcal{E}^\#$ be a product lifting from
$\mathcal{A}_1$ to $\mathcal{A}_2 \ot \mathcal{A}_1$ and let
$$\psi \ :\ \mathcal{S}(\mathcal{A}_2 \ot \mathcal{A}_1) \
\longrightarrow\ \mathcal{S}(\mathcal{A}_2 \ot \mathcal{A}_1)\ ,
$$ be a positive map. Define a new map $\varphi  :
\mathcal{S}(\mathcal{A}_1) \longrightarrow
\mathcal{S}(\mathcal{A}_1)$ {\em via}
\begin{equation}\label{}
    \varphi(\rho) := {\rm Tr}_2 [ \psi( \mathcal{E}^\#(
    \rho))] = {\rm Tr}_2 [ \psi( \omega \ot
    \rho)] \ ,
\end{equation}
where $\omega$ is a fixed state in $\mathcal{A}_2$. By construction
$\varphi$ is a linear positive map. Note, however, that if $\psi$ is
only positive but not completely positive then $\varphi$ needs not
be completely positive. As an example consider $\mathcal{A}_1 =
\mathcal{A}_2 = M_2(\mathbb{C})$ and let $\psi : M_4(\mathbb{C})
\longrightarrow M_4(\mathbb{C})$ be the Robertson map defined as
follows \cite{P2,P5}
\begin{equation}\label{psi}
    \psi\left( \begin{array}{c|c} X_{11} & X_{12} \\ \hline X_{21} & X_{22} \end{array}
\right) = \frac 12 \left( \begin{array}{c|c} \mathbb{I}_2\,
\mbox{Tr} X_{22} &  X_{12} + R_2(X_{21}) \\ \hline  X_{21} +
R_2(X_{12}) & \mathbb{I}_2\, \mbox{Tr} X_{11}
\end{array} \right) \ ,
\end{equation}
where we represented an element $X \in M_4(\mathbb{C})$ in the block
form $X = \sum_{i,j=1}^2 e_{ij} \ot X_{ij}$ with $X_{ij} \in
M_2(\mathbb{C})$. Finally, $R_2$ denotes the reduction map $R_2 :
M_2(\mathbb{C}) \longrightarrow M_2(\mathbb{C})$ defined by
\begin{equation}\label{}
    R_2(X) = \mathbb{I}_2 {\rm Tr}\, X - X \ ,
\end{equation}
which is known to positive but not completely positive. Now, if
$\mathcal{E}^\#$ is a product lifting $\mathcal{E}^\#(\rho) = \omega
\ot \rho$, then one finds
\begin{equation}\label{}
\varphi(\rho) = \frac 12 \left( \begin{array}{cc} 2\,\rho_{22} &
\rho_{12} + \rho_{21} \\ \rho_{12} + \rho_{21} & 2\,\rho_{11}
\end{array} \right) \ .
\end{equation}
Interestingly, the positive map $\varphi$ does not depend upon the
state $\omega$. To show that $\varphi$ is not completely positive
let us observe that
\begin{equation}\label{}
({\rm id} \ot \varphi)P^+_2 = \frac 14 \left( \begin{array}{cc|cc} 0
& 0 & 0 & 1 \\ 0 & 2 & 1 & 0
\\ \hline 0 & 1 & 2 & 0 \\ 1 & 0 & 0 & 0  \end{array} \right)\ ,
\end{equation}
is not positively defined ($P_2^+$ denotes the maximally entangled
state in $\mathbb{C}^2 \ot \mathbb{C}^2$, i.e. $P_2^+ = \frac 12
\sum_{i,j=1}^2 e_{ij} \ot e_{ij}$). }
\end{remark}

\subsection{Nonlinear liftings}

Let $\Lambda$ be a linear completely positive unital map
\begin{equation}
\Lambda\ : \ \mathcal{B}(\mathcal{H}) \ \longrightarrow \ \mathcal{B}(%
\mathcal{H})\ .
\end{equation}

\begin{definition}[\protect\cite{K-pi}]
A positive operator
\begin{equation}
\pi := \sum_{i,j=1}^d e_{ij} {\,\otimes\,} \Lambda(e_{ij})\ ,
\end{equation}
living in $\mathcal{H}_2 {\,\otimes\,} \mathcal{H}_1$
($\mathcal{H}_k = \mathbb{C}^d$) is called the quantum conditional
probability (QCP) operator.
\end{definition}
QCP generalizes classical conditional probability for the quantum
noncommutative case.\footnote{%
For slightly different approach to quantum conditional probability see \cite%
{Cerf-Adami} and \cite{AC}.} Note, that in the classical case one may define
\begin{equation}
\pi :=\sum_{i=1}^{d}e_{ii}{\,\otimes \,}\Lambda (e_{ii})\ ,
\end{equation}%
with
\begin{equation}
\Lambda (e_{ii})=\sum_{j=1}^{d}\,p_{i|j}\,e_{jj}\ .
\end{equation}%
Then
\begin{equation}
\sum_{i}\,p_{i|j}=1\ ,  \label{class-pi}
\end{equation}%
is equivalent to
\begin{equation}
\mbox{Tr}_{2}\,\pi =\mathbb{I}\ ,  \label{tr-pi-2}
\end{equation}%
which follows from $\Lambda (\mathbb{I})=\mathbb{I}$. Following
\cite{K-pi} let us define a nonlinear lifting by
\begin{equation}  \label{Phi-pi}
\mathcal{E}^\#(\rho) := (\mathbb{I} {\,\otimes\,} \rho^{\frac 12})
\, \pi\, (\mathbb{I} {\,\otimes\,} \rho^{\frac 12})\ .
\end{equation}

\begin{proposition}
A lifting $\mathcal{E}^\#$ satisfies
\begin{equation}
\mathrm{Tr}_2\, \mathcal{E}^\#(\rho) = \rho\ ,
\end{equation}
and
\begin{equation}
\mathrm{Tr}_1\, \mathcal{E}^\#(\rho) = \Lambda^\#(\rho)^{\mathrm{T}%
}\ .
\end{equation}
\end{proposition}

\begin{Remark}
\emph{Note, that in the special case if
\begin{equation}
\Lambda(x) = \mathbb{\mathbb{I}} \, \mathrm{Tr}\, \omega^T x \ ,
\end{equation}
for some density operator $\omega$ in $\mathcal{H}_2$, then (\ref{Phi-pi})
defines a linear product lifting
\begin{equation}
\mathcal{E}^\#(\rho) = \omega {\,\otimes\,} \rho\ .
\end{equation}
}
\end{Remark}

\begin{example}
\emph{\ Let $\Lambda (a)=UaU^{\ast }$, with unitary $U$. One has for the
lifting
\begin{eqnarray}
\mathcal{E}^{\# }(\rho ) &=&(\mathbb{I}{\,\otimes \,}\rho ^{\frac{1}{2}%
})\,\sum_{i,j=1}^{n}e_{ij}{\,\otimes \,}Ue_{ij}U^{\ast }\,(\mathbb{I}{%
\,\otimes \,}\rho ^{\frac{1}{2}})  \notag \\
&=&\sum_{i,j=1}^{n}e_{ij}{\,\otimes \,}\rho ^{\frac{1}{2}}Ue_{ij}U^{\ast
}\rho ^{\frac{1}{2}}\ ,
\end{eqnarray}%
that is
\begin{equation}
\mathcal{E}^{\# }(\rho )=\sum_{i,j=1}^{n}e_{ij}{\,\otimes \,}\phi
(e_{ij})\ ,
\end{equation}%
where $\phi $ is the $\rho $-dependent CP map
\begin{equation*}
\phi (a)=KaK^{\ast }\ ,
\end{equation*}%
with the $\rho $-dependent operator given by $K=\rho ^{\frac{1}{2}}U$. }
\end{example}

\begin{Example}
\emph{Let $\Lambda(a) = \sum_i \mbox{Tr}(a e_{ii})\, e_{ii}$. It leads to
the following lifting
\begin{eqnarray}
\mathcal{E}^\#(\rho) = \sum_{i=1}^n e_{ii} {\,\otimes\,} \rho^{\frac
12}\, e_{ii}\,\rho^{\frac 12}\ .
\end{eqnarray}
Note, that if $e_i$ defines eigen-basis for $\rho$, then
\begin{equation}
\mathcal{E}^\#(\rho) = \sum_{i=1}^n \rho_i\,e_{ii} {\,\otimes\,}
e_{ii}\ ,
\end{equation}
with $\rho_i$ being an eigenvalue of $\rho$, i.e. $\rho\, e_i = \rho_i\, e_i$%
. The above formula reproduces the nonlinear Ohya lifting.  Recall,
that the nonlinear Ohya lifting $\mathcal{E}^\# :
\mathcal{S}(\mathcal{A}) \longrightarrow \mathcal{S}(\mathcal{A}\ot
\mathcal{A})$ is defined as follows \cite{AO}
\begin{equation}\label{}
    \mathcal{E}^\#(\rho) = \sum_k p_k E_k \ot E_k \ ,
\end{equation}
where
\begin{equation}\label{}
    \rho = \sum_k p_k E_k\ ,
\end{equation}
stands for the spectral decomposition of $\rho$. We stress that
$\mathcal{E}^\#$ is nonlinear since both $p_k$ and $E_k$ are
$\rho$-dependent. }
\end{Example}

\begin{Remark}
\emph{Note, that a characteristic feature of the Ohya lifting is that both
reduced states reproduce the original state $\rho$, i.e.
\begin{equation}
\mathrm{Tr}_1\,\mathcal{E}^\#(\rho) = \mathrm{Tr}_2\,\mathcal{E}%
^\#(\rho) = \rho\ ,
\end{equation}
for an arbitrary classical state $\rho$ on $\Omega$. It shows that Ohya
lifting realizes a perfect classical cloning machine \cite{clon}. Moreover,
since both reduced states are the same Ohya lifting provides at the same
time a protocol for a classical broadcasting \cite{Ariano}. }
\end{Remark}

\begin{Remark}
\emph{Let $\theta$ be a compound state in $\mathcal{H}_2 {\,\otimes\,}
\mathcal{H}_1$ with marginals $\rho$ and $\sigma$. If $\phi$ is a CP map
such that
\begin{equation}
\theta = \sum_{i,j=1}^n e_{ij} {\,\otimes\,} \phi(e_{ij})\ ,
\end{equation}
then one can rewrite $\theta$ as follows
\begin{equation}
\theta = \sum_{i,j=1}^n e_{ij} {\,\otimes\,} \rho^{\frac
12}\Lambda(e_{ij})\rho^{\frac 12}\ ,
\end{equation}
where $\Lambda$ is a unital CP map defined by
\begin{equation}
\Lambda(a) = \rho^{-\frac 12}\phi(a) \rho^{-\frac 12}\ ,
\end{equation}
where we assume that $\rho$ is faithful state, i.e. $\rho > 0$.}
\end{Remark}

Suppose now that we have two QCP operators $\pi_1$ and $\pi_2$. Let us
define a 3-partite operator $\pi_1 \circ \pi_2$
\begin{equation}
\pi_1 \circ \pi_2 \ : \ \mathcal{H}_3 {\,\otimes\,} \mathcal{H}_2 {%
\,\otimes\,} \mathcal{H}_1 \ \longrightarrow \ \mathcal{H}_3 {\,\otimes\,}
\mathcal{H}_2 {\,\otimes\,} \mathcal{H}_1 \ ,
\end{equation}
with $\mathcal{H}_k = \mathbb{C}^d$, by
\begin{equation}
\pi_1 \circ \pi_2 := (\mathbb{I} {\,\otimes\,} \pi_1^{\frac 12})\, (\pi_2 {%
\,\otimes\,} \mathbb{I})\, (\mathbb{I} {\,\otimes\,} \pi_1^{\frac 12}) \ .
\end{equation}

\begin{proposition}
The operator $\pi _{1}\circ \pi _{2}$ is positive and it satisfies the
following basic properties:
\begin{eqnarray}
\mathrm{Tr}_{3}\,(\pi _{1}\circ \pi _{2}) &=&\pi _{1}\ ,  \label{pi-pi-1} \\
\mathrm{Tr}_{32}\,(\pi _{1}\circ \pi _{2}) &=&\mathbb{I}\ .
\end{eqnarray}
\end{proposition}

It is clear that for $\pi _{1}=\pi _{2}=\pi $, the operator $\pi \circ \pi $
may be used to define a lifting from $M_{d}^{+}$ into $M_{d}^{+}{\,\otimes \,%
}M_{d}^{+}{\,\otimes \,}M_{d}^{+}$. One defines
\begin{equation}
\mathcal{E}_{3|1}^{\# }(\rho )=(\mathbb{I}{\,\otimes \,}\mathbb{I}{%
\,\otimes \,}\rho ^{\frac{1}{2}})\,(\pi \circ \pi )\,(\mathbb{I}{\,\otimes \,%
}\mathbb{I}{\,\otimes \,}\rho ^{\frac{1}{2}})\ .
\end{equation}%
Note, that due to (\ref{pi-pi-1}) one has
\begin{equation}
\mbox{Tr}_{3}\,\mathcal{E}_{3|1}^{\# }(\rho )=\mathcal{E}_{2|1}^{\#
}(\rho )\ ,
\end{equation}%
where $\mathcal{E}_{2|1}^{\# }(\rho ):=\mathcal{E}^{\# }(\rho )$ is
defined in (\ref{Phi-pi}). This procedure may be immediately
generalized for arbitrary $N$. One has the following recurrence
formula for the $N$-partite operator
\begin{equation}
\pi _{1}\circ \ldots \circ \pi _{N-1}\ :\ \mathcal{H}_{N}{\,\otimes \,}%
\ldots {\,\otimes \,}\mathcal{H}_{1}\ \longrightarrow \ \mathcal{H}_{N}{%
\,\otimes \,}\ldots {\,\otimes \,}\mathcal{H}_{1}\ ,
\end{equation}%
\begin{equation}
\pi _{1}\circ \ldots \circ \pi _{N-1}:=(\mathbb{I}{\,\otimes \,}\ldots {%
\,\otimes \,}\mathbb{I}{\,\otimes \,}\pi _{1}^{\frac{1}{2}})\,(\pi _{2}\circ
\ldots \circ \pi _{N-1}{\,\otimes \,}\mathbb{I})\,(\mathbb{I}{\,\otimes \,}%
\ldots {\,\otimes \,}\mathbb{I}{\,\otimes \,}\pi _{1}^{\frac{1}{2}})\ .
\end{equation}%
One easily proves
\begin{eqnarray}
\mbox{Tr}_{N}\,(\pi _{1}\circ \ldots \circ \pi _{N-1}) &=&\pi _{1}\circ
\ldots \circ \pi _{N-2}\ ,  \label{pi-pi-2} \\
\mbox{Tr}_{N,N-1}\,(\pi _{1}\circ \ldots \circ \pi _{N-1}) &=&\pi _{1}\circ
\ldots \circ \pi _{N-3}\ , \\
&\vdots &  \notag \\
\mbox{Tr}_{N,\ldots ,2}\,(\pi _{1}\circ \ldots \circ \pi _{N-1}) &=&\mathbb{I%
}\ .
\end{eqnarray}%
Again, when $\pi _{1}=\ldots =\pi _{N-1}=\pi $, then $N$-partite $\pi \circ
\ldots \circ \pi $ is positive and one defines
\begin{equation}
\mathcal{E}_{N|1}^{\# }(\rho )=(\mathbb{I}{\,\otimes \,}\ldots {%
\,\otimes \,}\mathbb{I}{\,\otimes \,}\rho ^{\frac{1}{2}})\,(\pi \circ \ldots
\circ \pi )\,(\mathbb{I}{\,\otimes \,}\ldots {\,\otimes \,}\mathbb{I}{%
\,\otimes \,}\rho ^{\frac{1}{2}})\ .
\end{equation}%
It is clear that
\begin{equation}
\mbox{Tr}_{N}\,\mathcal{E}_{N|1}^{\# }(\rho )=\mathcal{E}%
_{N-1|1}^{\# }(\rho )\ .
\end{equation}%
Note, that in the classical case defined one finds
\begin{equation}
\mathcal{E}_{N|1}^{\# }(\rho )=\sum_{i_{1},\ldots
,i_{N}=1}^{d}\,p_{i_{N}|i_{N-1}}\ldots
p_{i_{2}|i_{1}}\,p_{i_{1}}\,e_{i_{N}i_{N}}{\,\otimes \,}\ldots {\,\otimes \,}%
e_{i_{1}i_{1}}\ ,
\end{equation}%
where $\rho =\sum_{i}p_{i}e_{ii}$. Hence, the lifted $N$-partite state $%
\mathcal{E}^{\# }(\rho )$ generalizes classical Markovian state (for
entanglement properties of {a particular class of pure Markovian}
multipartite state see \cite{AMO}).

\section{Circulant liftings}

\label{CIRC} \setcounter{equation}{0}

In this section we analyze a particular class of liftings defined in
terms of circulant states \cite{CIRCULANT} (see also \cite{ART}).
Circulant states play important role in Quantum Information Theory
since majority of states of composite quantum systems considered in
the literature turn out to be circulant states. The most important
example of circulant states is provided by generalized Bell  states.
Let us mention also the class of isotropic states, Werner states,
Bell diagonal states and many others (see \cite{CIRCULANT} for more
examples). Therefore, it is interesting to investigate the class of
liftings directly related to the class of circulant states. This
section provides also the illustration of the theoretical concepts
introduced so far. It is organized as follows: we start with the
notion of circulant decomposition of any Hilbert space being the
tensor product $\mathbb{C}^d \ot \mathbb{C}^d$. It turns out that it
may be represented as a direct product of $d$ subspaces each of
dimension $d$. This decomposition (we call it {\em cyclic
decomposition}) provides starting point in the construction of a
circulant state. Having defined circulant state we analyze the class
of liftings such that $\mathcal{E}^\#(\rho)$ is a circulant state
for any $\rho$ living in $\mathbb{C}^d$ and illustrate the
construction by the special class of Bell diagonal liftings.

\subsection{Circulant decompositions}

The basic idea is to decompose the total Hilbert space $\mathbb{C}^d {%
\,\otimes\,} \mathbb{C}^d$ into a direct sum of $d$ orthogonal $d$%
-dimensional subspaces related by a certain cyclic property. Let us
introduce a $d$-dimensional subspace
\begin{equation}
{\Sigma}_0 = \mbox{span}\left\{ e_0 {\,\otimes\,} e_0\, , e_1 {\,\otimes\,}
e_1\, , \ldots\, , e_{d-1} {\,\otimes\,} e_{d-1} \right\} \ ,
\end{equation}
where $\{e_0,\ldots,e_{d-1}\}$ denote the standard computational basis in $%
\mathbb{C}^d$, together with the shift operator $S : \mathbb{C}^d
\longrightarrow \mathbb{C}^d$ defined by
\begin{equation}  \label{S}
Se_k = e_{k+1} \ , \ \ \ \mathrm{mod} \ d\ .
\end{equation}
Now, for any $\alpha=0,\ldots,d-1$, we define
\begin{equation}
{\Sigma}_\alpha = ({\mathchoice{\rm 1\mskip-4mu l}{\rm 1\mskip-4mu l}{\rm
1\mskip-4.5mu l}{\rm 1\mskip-5mu l}} {\,\otimes\,} S^\alpha) {\Sigma}_0\ .
\end{equation}
It is clear that $\Sigma_\alpha$ and $\Sigma_\beta$ are mutually orthogonal
(for $\alpha\neq \beta$). Moreover,
\begin{equation}  \label{D}
\Sigma_0 \oplus \ldots \oplus \Sigma_{d-1} = \mathbb{C}^d {\,\otimes\,}
\mathbb{C}^d \ .
\end{equation}
Due to the cyclic property of the shift operator (\ref{S}), we call (\ref{D}%
) a circulant decomposition \cite{CIRCULANT}. For example in the case of two
qubits ($d=2)$ one has
\begin{eqnarray}
\Sigma_0 &=& \mbox{span}\left\{ e_0 {\,\otimes\,} e_0\, , e_1 {\,\otimes\,}
e_1 \right\} \ ,  \notag \\
\Sigma_1 &=& \mbox{span}\left\{ e_0 {\,\otimes\,} e_1\, , e_1 {\,\otimes\,}
e_0 \right\} \ ,
\end{eqnarray}
whereas for two qutrits ($d=3$) one obtains
\begin{eqnarray}
\Sigma_0 &=& \mbox{span}\left\{ e_0 {\,\otimes\,} e_0\, , e_1 {\,\otimes\,}
e_1\, , e_2{\,\otimes\,} e_2 \right\} \ ,  \notag \\
\Sigma_1 &=& \mbox{span}\left\{ e_0 {\,\otimes\,} e_1\, , e_1 {\,\otimes\,}
e_2\, , e_2{\,\otimes\,} e_0 \right\} \ ,  \notag \\
\Sigma_2 &=& \mbox{span}\left\{ e_0 {\,\otimes\,} e_2\, , e_1 {\,\otimes\,}
e_0\, , e_2{\,\otimes\,} e_1 \right\} \ .
\end{eqnarray}

\subsection{Circulant states}

{The} circulant decomposition (\ref{D}) gives rise to the following
construction of {a} state in $\mathbb{C}^{d}{\,\otimes \,}\mathbb{C}%
^{d}$: let $a^{(0)},\ldots ,a^{(d-1)}$ be a set of $d$ positive $d\times d$
matrices. Let us observe that
\begin{equation}
\rho ^{(\alpha )}=\sum_{i,j=0}^{d-1}a_{ij}^{(\alpha )}\ e_{ij}{\,\otimes \,}%
S^{\alpha }e_{ij}S^{\ast \alpha }\ ,\ \ \ \ \ \alpha =0,\ldots ,d-1\ ,
\end{equation}%
is a positive operator in $\mathbb{C}^{d}{\,\otimes \,}\mathbb{C}^{d}$
supported on $\Sigma _{\alpha }$. Hence, the following sum
\begin{equation}
\rho =\rho ^{(0)}+\ldots +\rho ^{(d-1)}\ ,  \label{ro-C}
\end{equation}%
gives rise to a positive operator in $\mathbb{C}^{d}{\,\otimes \,}\mathbb{C}%
^{d}$. It defines a legitimate state iff $\mbox{Tr}\rho =1$, which is
equivalent to
\begin{equation}
\mathrm{Tr}\left[ a^{(0)}+\ldots +a^{(d-1)}\right] =1\ .
\end{equation}%
For obvious reason we call (\ref{ro-C}) a circulant state. This simple
construction recovers many well known bipartite states from the literature
(see \cite{CIRCULANT}).

Consider now a partial transposition of the circulant state (\ref{ro-C}). It
turns out that $\rho^\Gamma = ({%
\mathchoice{\rm 1\mskip-4mu l}{\rm 1\mskip-4mu l}{\rm 1\mskip-4.5mu
l}{\rm 1\mskip-5mu l}} {\,\otimes\,} \tau)\rho$ is again circulant
but it corresponds to another cyclic decomposition of the original
Hilbert space $\mathbb{C}^d {\,\otimes\,} \mathbb{C}^d$. Recall,
that $\rho$ is PPT (Positive Partial Transpose) if $\rho^\Gamma \geq
0$. Let us introduce the following permutation $\pi$ from the
symmetric group $S_d$:
\begin{equation}
\pi(0) = 0 \ , \ \ \ \ \pi(i) = d-i \ , \ \ i=1,2,\ldots,d-1\ .
\end{equation}
We use $\pi$ to introduce
\begin{equation}
\widetilde{{\Sigma}}_0 = \mbox{span}\left\{ e_0 {\,\otimes\,} e_{\pi(0)}\, ,
e_1 {\,\otimes\,} e_{\pi(1)}\, , \ldots\, , e_{d-1} {\,\otimes\,}
e_{\pi(d-1)} \right\} \ ,
\end{equation}
and
\begin{equation}
\widetilde{{\Sigma}}_\alpha = ({\mathchoice{\rm 1\mskip-4mu l}{\rm
1\mskip-4mu l}{\rm 1\mskip-4.5mu l}{\rm 1\mskip-5mu l}} {\,\otimes\,}
S^\alpha) \widetilde{{\Sigma}}_0\ .
\end{equation}
It is clear that $\widetilde{\Sigma}_\alpha$ and $\widetilde{\Sigma}_\beta$
are mutually orthogonal (for $\alpha\neq \beta$). Moreover,
\begin{equation}  \label{D-new}
\widetilde{\Sigma}_0 \oplus \ldots \oplus \widetilde{\Sigma}_{d-1} = \mathbb{%
C}^d {\,\otimes\,} \mathbb{C}^d \ ,
\end{equation}
and hence it defines another circulant decomposition. Note, that for $d=2$
one has $\widetilde{\Sigma}_\alpha = {\Sigma}_\alpha$. It is no longer true
for $d\geq 3$. For $d=3$ one finds
\begin{eqnarray}
\widetilde{\Sigma}_0 &=& \mbox{span}\left\{ e_0 {\,\otimes\,} e_0\, , e_1 {%
\,\otimes\,} e_2\, , e_2 {\,\otimes\,} e_1 \right\} \ ,  \notag \\
\widetilde{\Sigma}_1 &=& \mbox{span}\left\{ e_0 {\,\otimes\,} e_1\, , e_1 {%
\,\otimes\,} e_0\, , e_2{\,\otimes\,} e_2 \right\} \ , \\
\widetilde{\Sigma}_2 &=& \mbox{span}\left\{ e_0 {\,\otimes\,} e_2\, , e_1 {%
\,\otimes\,} e_1\, , e_2{\,\otimes\,} e_0 \right\} \ .  \notag
\end{eqnarray}
Now, the partially transformed state $\rho^\tau$ has again a circulant
structure but with respect to the new decomposition (\ref{D-new}):
\begin{equation}  \label{ro-C-new}
\rho^\Gamma = \widetilde{\rho}^{(0)} + \ldots + \widetilde{\rho}^{(d-1)} \ ,
\end{equation}
where
\begin{equation}
\widetilde{\rho}^{(\alpha)} = \sum_{i,j=0}^{d-1} \widetilde{a}%
^{(\alpha)}_{ij} \ e_{ij} {\,\otimes\,} S^\alpha e_{\pi(i)\pi(j)} S^{\dag
\alpha} \ ,\ \ \ \ \ \alpha=0,\ldots,d-1 \ ,
\end{equation}
and the new $d \times d$ matrices $[\widetilde{a}^{(\alpha)}_{ij}]$ are
given by the following formulae:
\begin{equation}  \label{a-tilde}
\widetilde{a}^{(\alpha)} \, =\, \sum_{\beta=0}^{d-1}\, a^{(\alpha+\beta)}
\circ ({\Pi} {S}^\beta)\ , \ \ \ \ \ \ \ \ \mbox{mod $d$}\ ,
\end{equation}
where ``$\circ$" denotes the Hadamard product,\footnote{%
A Hadamard (or Schur) product of two $n \times n$ matrices $A=[A_{ij}]$ and $%
B=[B_{ij}]$ is defined by
\begin{equation*}
(A \circ B)_{ij} = A_{ij} B_{ij}\ .
\end{equation*}%
} and $\Pi$ being a $d \times d$ permutation matrix corresponding to $\pi$,
i.e. $\Pi_{ij} := \delta_{i,\pi(j)}$. It is therefore clear that our
original circulant state (\ref{ro-C}) is PPT iff all $d$ matrices $%
\widetilde{a}^{(\alpha)}$ satisfy
\begin{equation}
\widetilde{a}^{(\alpha)} \geq 0 \ , \ \ \ \ \alpha=0,\ldots,d-1\ .
\end{equation}

\subsection{Circulant liftings}

Circulant states provide interesting example of a linear lifting. Denote by $%
M_d$ a $\mathbb{C}^*$-algebra of $d \times d$ complex matrices and consider
the following lifting
\begin{equation}
\mathcal{E}^\# \ :\ S(M_d)\ \longrightarrow \ S(M_d {\,\otimes\,}
M_d) \ ,
\end{equation}
defined by
\begin{equation}
\mathcal{E}^\#({\rho}) = \sum_{\alpha=0}^{d-1}\, c^{(\alpha)}_{ij}
\, e_{ij} {\,\otimes\,} V_{i\alpha} \rho V_{i\alpha }^*\ ,
\end{equation}
where $[c^{(\alpha)}_{ij}]$ are $d \times d$ positive matrices for $%
\alpha=0,1,\ldots,d-1$ such that $\mathrm{Tr}\, c^{(\alpha)} =1 $, and
\begin{equation}
V_{i\alpha} = |e_i+e_\alpha\>\<e_\alpha| \ .
\end{equation}
Note that
\begin{equation}
\mathcal{E}^\#({\rho}) = \sum_{\alpha=0}^{d-1}\ a^{(\alpha)}_{ij} \,
e_{ij} {\,\otimes\,} S^\alpha e_{ij} S^{\alpha *}\ ,
\end{equation}
where
\begin{equation}
a^{(\alpha)}_{ij} = p_\alpha\, c^{(\alpha)}_{ij} \ ,
\end{equation}
and
\begin{equation}  \label{p-alpha}
p_\alpha = \rho_{\alpha\alpha} \ .
\end{equation}
It shows that for any $\rho$ its lifting $\mathcal{E}^\#({\rho})$
defines a circulant state. Now, if $c^{(\alpha)}$ are rank-1
projectors, i.e. ${c}^{(\alpha)}_{mn} = c^{(\alpha)}_m\,
\overline{c}^{(\alpha)}_n$ are the Grahm matrices for the $d$
complex $d$-vectors $c^{(\alpha)}$, the above formula simplifies to
\begin{equation}  \label{E-pure}
\mathcal{E}^\#(\rho)\ := \ V \, \mathcal{D}(\rho)\, V^*\ ,
\end{equation}
where
\begin{equation}
\mathcal{D}(\rho) := \sum_{i=0}^{d-1}\, e_{ii}\,\rho \,e_{ii}\ ,
\end{equation}
is the projection onto the diagonal part of $\rho$, and
\begin{equation}
V \ :\ \mathbb{C}^d\ \longrightarrow \ \mathbb{C}^d {\,\otimes\,} \mathbb{C}%
^d
\end{equation}
is defined by
\begin{equation}
V e_\alpha\ =\ \sum_{j=0}^{d-1} c^{(\alpha)}_j\, e_j {\,\otimes\,}
e_{j+\alpha} \ {\ }
\end{equation}
Note, that due the trace condition $\mathrm{Tr} c^{(\alpha)} =1 $ the linear
operator $V$ defines an isometry
\begin{equation}
V^* V = \mathbb{I}\ .
\end{equation}
It should be stressed that the above circulant lifting is never pure.

\subsection{Bell diagonal lifting}

Consider a simplex of Bell diagonal states \cite{Bell-1,Bell-2} defined by
\begin{equation}  \label{Bell}
\rho = \sum_{m,n=0}^{d-1} p_{mn} P_{mn}\ ,
\end{equation}
where $p_{mn}\geq 0$, $\ \sum_{m,n}p_{mn}=1$ and
\begin{equation}
P_{mn} = (\mathbb{I} \ot U_{mn}) P^+_d (\mathbb{I}\ot
U_{mn}^\dagger)\ ,
\end{equation}
with $U_{mn}$ being the collection of $d^2$ unitary matrices defined as
follows
\begin{equation}  \label{U_mn}
U_{mn} e_k = \lambda^{mk} S^n e_k = \lambda^{mk} e_{k+n}\ ,
\end{equation}
with
\begin{equation}
\lambda= e^{2\pi i/d} \ .
\end{equation}
The matrices $U_{mn}$ define an orthonormal basis in the space $M_d(\mathbb{C%
})$ of complex $d \times d$ matrices. One easily shows
\begin{equation}
\mathrm{Tr}(U_{mn} U_{rs}^\dagger) = d\, \delta_{mr} \delta_{ns} \ .
\end{equation}
It is clear \cite{Taka} that Bell diagonal states define a subclass of
circulant states.

\begin{definition}
A circulant lifting $\mathcal{E}^\#$ is Bell diagonal if $\mathcal{E}%
^\#(\rho)$ defines Bell diagonal state for any $\rho$.
\end{definition}
Suppose now that $c^{(0)} = \ldots = c^{(d-1)} =: c$, where the positive
matrix $c$ is defined by
\begin{equation}
c_{kl} = \frac 1d \sum_{m=0}^{d-1} p_m \lambda^{m(k-l)}\ .
\end{equation}
The above formula defines a circulant matrix\footnote{Recall,  that
a $d \times d$ matrix $c_{ij}$ is circulant if $c_{ij}$ depends only
upon the difference $i-j$ (mod $d$). For example
\[ \left( \begin{array}{cc} a & b \\ b & a \end{array} \right) \ , \ \ \ \
\left( \begin{array}{ccc} a & b & c \\ c & a & b \\ b & c & a
\end{array} \right)\ ,
\]
are circulant matrices.}
\begin{equation}
c_{kl} = f_{k-l}\ ,
\end{equation}
where the $d$-vector $f_n$ is defined via the discrete Fourier transform of
the probability $d$-vector $p_m$. One finds for the lifted state $\rho$
\begin{equation}
\mathcal{E}^\# (\rho) = \sum_{m,n=0}^{d-1} p_{mn}(\rho) P_{mn} \ ,
\end{equation}
where
\begin{equation}
p_{mn}(\rho) = p_m \<n|\rho|n\> \ .
\end{equation}
Hence, the joint distribution $p_{mn}$ is the product of $p_m$ and the
classical probability distribution defined by the diagonal elements $%
\rho_{nn}$ of $\rho$. For more detailed analysis of Bell diagonal
states we refer to the recent paper \cite{Taka}.

\section{Conclusions}

We analyzed the procedure of lifting in classical stochastic and
quantum systems. Lifting enables one to `lift' a state of a
classical/quantum system into a state of `system+reservoir'. This
procedure is important both in quantum information theory and {in}
the theory of open systems. It turns out that it is very much
related to the problem of completely positive maps are the workhorse
in these fields. We discussed both linear and nonlinear liftings and
provided instructive illustration of the general theory by a
particular class related to so called circulant states. In
particular it is shown {how} to lift a state of a subsystem to the
Bell diagonal state of the composed system. The theory of liftings
may provide new constructions of classical/quantum channels.
Moreover, it may be used to construct new classes of linear maps
which are positive but not completely positive. It is well {known}
that such maps define a basic tool for detection quantum
entanglement. We therefore conclude that the theory of liftings
might provide an interesting insight both in quantum information
theory and \textbf{in} the intricate mathematics of (completely)
positive maps. Both problems deserve further studies.

\section*{Acknowledgement}

DC and AK were supported by the Polish Ministry of Science and
Higher Education Grant No 3004/B/H03/2007/33. {TM was supported by
the QBIC grant.}

\end{document}